# Friedel oscillations in 2D electron gas from spin-orbit interaction in a parallel magnetic field

I. V. Kozlov and Yu. A. Kolesnichenko[a)]

*B.I. Verkin Institute for Low Temperature Physics and Engineering, National Academy of Sciences of Ukraine, pr. Nauki, 47, Khar'kov, 61103, Ukraine*

Effects associated with the interference of electron waves around a magnetic point defect in two-dimensional electron gas with combined Rashba-Dresselhaus spin-orbit interaction in the presence of a parallel magnetic field are theoretically investigated. The effect of a magnetic field on the anisotropic spatial distribution of the local density of states and the local density of magnetization is analyzed. The existence of oscillations of the density of magnetization with scattering by a non-magnetic defect and the contribution of magnetic scattering (accompanied by spin-flip) in the local density of electron states are predicted.

## 1. Introduction

Interest in research on quasi-two-dimensional [hereinafter for brevity, two-dimensional (2D)] conductive systems results from new capabilities for studying various quantum effects that are absent or are very small in mass conductors. Among such effects are spatial oscillations of the density of electrons $n(\epsilon_F, r)$ theoretically predicted by Friedel,[1] depending on the distance $r$ from a point defect or conductor edge ($\epsilon_F$ is Fermi energy). Friedel oscillations (FO) are associated with interference of incident and reflected electron waves and in an isotropic conductor have a period $\Delta r = \pi\hbar/p_F$ (where $p_F$ is the Fermi impulse). Direct observation of FO became possible with the development of scanning tunnel microscopy (STM).[2] Thus, when studying in Refs. [3] and [4] the surface (111) of noble metals by means of STM, FO of the local density of states (LDS) $\rho(\epsilon, r) = dn/d\epsilon|_{\epsilon = \epsilon_F}$ were detected, arising as the result of the scattering of 2D surface states on individual adsorbed atoms. It was further shown that the study of the periods and contours of the constant phase of FO makes it possible to obtain new information on the local characteristics of the spectrum of charge carriers and on the process of their scattering by an individual point defect of known nature (see reviews[5–7] and the references quoted in them).

During electron scattering on a magnetic impurity at temperature $T$ below the Kondo temperature $T_K$,[8] there appears in the FO of the electron density $n(\epsilon_F, r)$ an additional shift of the phase of the oscillations depending on $r/r_K$ ($r_K = \hbar v_F / T_K$ is the Kondo length) which, however, is absent in oscillations $\rho(\epsilon_F, r)$ of the LDS as measured by means of STM.[9] The magnetic defect, along with oscillations of the electron density, results in oscillations of the local density of magnetization (LDM) of electron gas $\mathbf{m}(\epsilon_F, r)$, which is a magnetic analogue of the FO[10] (hereinafter we will use for brevity the term FO of the LDM). Spatial oscillations of the LDM are investigated by means of a spin-polarized scanning tunnel microscope having a contact (tip) with a magnetic covering.[11]

The asymmetrical confinement electric potential (confinement potential) that limits the motion of electrons along the normal to the edge (for surface states) or to the border of the heterotransition with a quantum well, as well as the absence of a center of inversion of a bulk crystal, result in spin-orbit interaction (SOI) (see the monograph,[12] which significantly affects the thermodynamic and kinetic characteristics of such 2D systems.[13] The FO near an individual point defect in 2D electron gas with Rashba SOI[18,19] were experimentally observed in Refs. [14–17], and in this case a rather large number of theoretical works[20–27] were devoted to the analysis of the oscillatory dependence of the LDS. The LDM was theoretically studied in Ref. [28] for electron scattering of surface states of Au(111) on an adsorbed Co atom, taking into account the Rashba SOI and the Kondo effect.

In a number of 2D systems, combined Rashba[19] and Dresselhaus[31] SOI (R-D SOI) is produced, based on semiconductors such as zincblende (III – V zincblende and wurtzite) and SiGe (see Refs. [29] and [30]). With R-D SOI, the scattering law of charge carriers is anisotropic, and this results in significant modification of the FO. Thus, FO beats were predicted in Ref. [32], and it is shown in Ref. [33] that the FO are also essentially anisotropic, and contain more than two harmonics for a certain relationship of the constants of SOI.

As a consequence of the dependence on the wave vector of the direction of electron spin in 2D systems with SOI, the parallel magnetic field not only results in Zeeman splitting, but also significantly changes the energy spectrum (see for example Refs. [34] and [35]). By changing the amplitude and direction of the magnetic field, it is possible by a change in the spectrum to control all electron characteristics of a 2D conductor with R-D SOI.

In this article we studied FO oscillations around the point of a magnetic defect in 2D electron gas with R-D SOI located in a parallel magnetic field. General expressions for the LDS and LDM and their asymptotic expressions at large distances from the defect were obtained in the Born approximation. The dependences of the FO of the LDS and LDM on the amplitude and direction of the magnetic field were analyzed. The effect of the occurrence of oscillations of the density of magnetization during scattering by a non-magnetic defect and the dependence of the oscillations of the density of states on the magnitude of the magnetic moment of the defect were predicted.

## 2. Statement of the problem

### 2.1. The Hamiltonian

Using the calibration $\mathbf{A} = (0, 0, B_x y - B_y x)$, we write the Hamiltonian of the 2D electron gas with R-D SOI in a parallel magnetic field $\mathbf{B} = (B_x, B_y, 0)$ in the absence of defects, as a linear approximation on the wave vector operator $\hat{\mathbf{k}} = -i\nabla_{\mathbf{r}}$ (see, for example, Refs. 34 and 35):

$$\hat{H}_0 = \frac{\hbar^2(\hat{k}_x^2 + \hat{k}_y^2)}{2m}\sigma_0 + \alpha(\sigma_x \hat{k}_y - \sigma_y \hat{k}_x) + \beta(\sigma_x \hat{k}_x - \sigma_y \hat{k}_y) + \frac{g^*}{2}\mu_B(B_x\sigma_x + B_y\sigma_y), \quad (1)$$

where $m$ is the effective mass of an electron, $\mathbf{B} = (B_x, B_y, 0)$ is the induction of the magnetic field, $\sigma_{x,y,z}$ are Pauli matrices, $\hat{\sigma}_0$ is the 2 × 2 identity matrix, $\alpha$ and $\beta$ are constants of the Rashba ($\alpha$) and Dresselhaus ($\beta$) SOI, $\mu_B$ is the Bohr magneton, and $g^*$ is the effective g-factor of the 2D system, which can significantly differ from the value $g_0 = 2$ for free electrons.[12] For definiteness we will assume that the spin-orbit interaction constants are positive.

The eigenvalues and eigenfunctions of the Hamiltonian in (1) are written as

$$\epsilon_{1,2}(\mathbf{k}) = \frac{\hbar^2 k^2}{2m} \pm A(k_x, k_y),$$
$$A = \sqrt{(h_x + \alpha k_y + \beta k_x)^2 + (h_y + \alpha k_x + \beta k_y)^2}, \quad (2)$$

$$\psi_{1,2}(\mathbf{r}) = \frac{1}{2\pi\sqrt{2}}e^{i\mathbf{kr}}\begin{pmatrix}1 \\ e^{i\theta_{1,2}}\end{pmatrix} \equiv \frac{1}{2\pi}e^{i\mathbf{kr}}\phi(\theta_{1,2}), \quad (3)$$

where the $\phi(\theta_{1,2})$ are the spin parts of the wave functions of (3),

$$\sin\theta_1(\mathbf{k}) = \frac{h_y - \alpha k_x - \beta k_y}{A}, \quad \cos\theta_1(\mathbf{k})$$
$$= \frac{h_x - \alpha k_y - \beta k_x}{A}, \quad \theta_2 = \theta_1 + \pi. \quad (4)$$

In order to reduce writing in later formulas, we introduce the designation

$$\mathbf{h} = \frac{g^*}{2}\mu_B \mathbf{B}. \quad (5)$$

The spin orientation for each of the branches of the spectrum in (2) is defined by the mean

$$\mathbf{s}_{1,2}(\theta) = \phi^\dagger(\theta_{1,2})\sigma\phi(\theta_{1,2}) = (\cos\theta_{1,2}, \sin\theta_{1,2}, 0), \quad (6)$$

and, in accordance with the formulas in (4), this depends not only on the direction of the wave vector, but also on its value, for combined R-D SOI in the presence of a parallel magnetic field.

### 2.2. Scattering by a defect

We will model the interaction of electrons with a point magnetic defect at the point $\mathbf{r} = 0$ using the two-dimensional $\delta$-potential, often used in the examination of various physical problems:[36]

$$D(\mathbf{r}) = \left[\gamma\sigma_0 + \frac{1}{2}\mathbf{J}\sigma\right]\delta(\mathbf{r}), \quad (7)$$

where $\gamma > 0$ is the constant of potential interaction of electrons with the defect; $\hat{\sigma} = (\hat{\sigma}_x, \hat{\sigma}_y, \hat{\sigma}_z)$ is the Pauli vector; and $\mathbf{J}$ is the effective magnetic moment of the defect which differs from its true value $\mathbf{S}$ ($S \geq 1$) due to the Kondo effect, resulting in partial screening of the spin $\mathbf{S}$ by conductivity electrons. We consider the direction of the vector $\mathbf{J}$ to be fixed, and will not consider the processes of revolution and precession of the defect spin.

Temperature $T$ is assumed equal to zero. Such an approach is quite justifiable since the quantum interferential phenomena to which FO pertain are usually observed at low temperatures, when scattering of electrons on phonons is sufficiently small. For $T = 0$, the LDS $\rho(\epsilon_F, \mathbf{r})$ and LDM $\mathbf{m}(\epsilon_F, \mathbf{r})$ can be calculated with the aid of the retarded Green function $\hat{G}^R(E, \mathbf{r}_1, \mathbf{r}_2)$ in the coordinate representation

$$\rho(\epsilon_F, \mathbf{r}) = -\frac{1}{\pi}\text{Im Sp}\left[\hat{G}^R(\epsilon_F, \mathbf{r}, \mathbf{r})\right], \quad (8)$$

$$\mathbf{m}(\epsilon_F, \mathbf{r}) = -\frac{1}{\pi}\text{Im Sp}\left[\hat{\sigma}\hat{G}^R(\epsilon_F, \mathbf{r}, \mathbf{r})\right]. \quad (9)$$

We will take account of the effect of electron scattering at a defect in the Born approximation (see, for example, Ref. 37 from the potential of scattering (7), after presenting the Green function in the form of a decomposition

$$\hat{G}^R(\epsilon, \mathbf{r}_1, \mathbf{r}_2) \approx \hat{G}_0^R(\epsilon, \mathbf{r}_1 - \mathbf{r}_2) + \hat{G}_0^R(\epsilon, \mathbf{r}_1)D(\mathbf{r})\hat{G}_0^R(\epsilon, -\mathbf{r}_2), \quad (10)$$

in which the retarded Green function in the absence of defects, $\hat{G}_0^R(\epsilon_F, \mathbf{r}_1 - \mathbf{r}_2)$, depends only on the difference of coordinates $\mathbf{r} = \mathbf{r}_1 - \mathbf{r}_2$:

$$\hat{G}_0^R(\epsilon, \mathbf{r}) = \frac{1}{(2\pi)^2}\int_{-\infty}^{\infty}\frac{d^2 k e^{i\mathbf{kr}}}{\epsilon - \hat{H}_0 + i0}; \quad \epsilon \in \mathbb{R}. \quad (11)$$

The Hamiltonian $\hat{H}_0$ is defined by expression (1). Naturally, formula (10) can be used to describe the FO only at sufficiently large distances from the defect $r > r_D$ when the term associated with scattering is small. The value of $r_D$ with respect to the order of magnitude of size can be estimated as the Fermi wavelength $r_D \sim \lambda_F \sim \hbar/p_F$ for potential scattering, and as the Kondo length $r_D - r_{K} = \hbar v_F/T_K$ for magnetic scattering.

## 3. The Green function

### 3.1. General ratios

We derived in Ref. 38 exact analytical expressions for the equilibrium Green function at temperature zero for a 2D system of electrons with R-D SOI, and their asymptotes for large values of the spatial variable. Here we will provide certain relationships based on earlier-obtained results[38] which will be necessary for further calculations

The equilibrium retarded Green function (11) can be presented in the form of a decomposition on the Pauli

matrices

$$\hat{G}_0^R(\epsilon, \mathbf{r}) = g_0(\epsilon + i0, \mathbf{r})\hat{\sigma}_0 + g_x(\epsilon + i0, \mathbf{r})\hat{\sigma}_x + g_y(\epsilon + i0, \mathbf{r})\hat{\sigma}_y, \epsilon, \in \mathbb{R}, \quad (12)$$

where

$$g_0(\epsilon, \mathbf{r}) = \frac{1}{2(2\pi)^2} \sum_{j=1,2} \int d^2k e^{i\mathbf{k}\mathbf{r}} \frac{1}{\epsilon - \epsilon_j(\mathbf{k})}, \quad (13)$$

$$\left\{ \begin{array}{c} g_x(\epsilon, \mathbf{r}) \\ g_y(\epsilon, \mathbf{r}) \end{array} \right\} = \frac{1}{2(2\pi)^2} \sum_{j=1,2} \int d^2k e^{i\mathbf{k}\mathbf{r}} \left\{ \begin{array}{c} \cos\theta_j(\mathbf{k}) \\ \sin\theta_j(\mathbf{k}) \end{array} \right\} \frac{1}{\epsilon - \epsilon_j(\mathbf{k})}, \quad (14)$$

$\theta_j$ is the angle defining the direction of spin of an electron (6). Its dependence on the pulse is defined by formula (3). Each of the terms with $j = 1, 2$ contributes to the Green function of one of the branches of the energy spectrum $\epsilon_{1,2}(2)$. The components $g_{0, x, y}$ of the Green function (12) satisfy the symmetry relationships

$$g_0(\mathbf{r}, \mathbf{h}) = g_0(-\mathbf{r}, -\mathbf{h}), \ g_{x,y}(\mathbf{r}, \mathbf{h}) = -g_{x,y}(-\mathbf{r}, -\mathbf{h}). \quad (15)$$

Further, we will assume $\alpha \neq \beta$ in all calculations, except for special cases when the equality of the constants of SOI is stipulated separately. For $\alpha \neq \beta$, it is convenient to transition to new variables of integration $\hat{k}$ and $f$.

$$k_x = k_{x0} + \tilde{k}\cos f, \ k_y = k_{y0} + \tilde{k}\sin f, \quad (16)$$

where

$$k_{x0} = h\frac{\alpha\sin\varphi_h + \beta\cos\varphi_h}{\alpha^2 - \beta^2}; \ k_{y0} = -h\frac{\alpha\cos\varphi_h + \beta\sin\varphi_h}{\alpha^2 - \beta^2} \quad (17)$$

are the coordinates of the point of contact of the branches of the spectrum, in which the energy is

$$\epsilon_{1,2}(\mathbf{k}_0) = \frac{\hbar^2(k_{0x}^2 + k_{0y}^2)}{2m} = h^2\hbar^2\frac{\alpha^2 + \beta^2 + 2\alpha\beta\sin 2\varphi_h}{2m(\alpha^2 - \beta^2)^2}$$
$$= \epsilon_0 > 0, \quad (18)$$

and the angle $\varphi_h$ specifies the direction of the magnetic field $\mathbf{h} = h(\cos\varphi_h, \sin\varphi_h, 0)$. The wave vector $\mathbf{k}_0 = (k_{x0}, k_{y0})$, corresponding to a point with energy $\epsilon_0$, is determined from the condition $A(k_{x0}, k_{y0}) = 0$ in the expressions in (2).

In the variables in (16), the dependence of energy $\epsilon_{1, 2}$ on $\tilde{k}$ and $f$ is written as

$$\epsilon_{1,2}(\tilde{k}, f) = \frac{\hbar^2\tilde{k}^2}{2m} - \frac{\hbar^2\tilde{k}}{2}\lambda_{1,2}(f) + \epsilon_0, \quad (19)$$

where

$$\lambda^{(1,2)}(f) = h\frac{\alpha\sin(f - \varphi_h) - \beta\cos(f + \varphi_h)}{\alpha^2 - \beta^2}$$
$$\mp \frac{m}{\hbar^2}\sqrt{\alpha^2 + \beta^2 + 2\alpha\beta\sin(2f)}. \quad (20)$$

In the new coordinates of (16), the point of contact of the branches of the spectrum corresponds to $\hat{k} = 0$.

The angles defining the direction of spin, $\theta_{1, 2}(f)$, depend only on the direction of the wave vector (angle $f$) and the constants of SOI, after the replacement in (16):

$$\sin\theta_{1,2}(f) = \mp \frac{\alpha\cos f + \beta\sin f}{\sqrt{\alpha^2 + \beta^2 + 2\alpha\beta\sin 2f}},$$

$$\cos\theta_{1,2}(f) = \pm \frac{\alpha\sin f + \beta\cos f}{\sqrt{\alpha^2 + \beta^2 + 2\alpha\beta\sin 2f}}. \quad (21)$$

Hence, for each branch of the spectrum, the directions of electron spin $\mathbf{s}_{1,2}(f)$ are antisymmetric relative to the point $\hat{k} = 0$. For each angle $f$, the directions of spin of electron belonging to different branches of the spectrum are strictly opposite.

Substituting expressions (16) and (21) into formulas (13) and (14), after integration on $\hat{k}$ we derive

$$g_0(\epsilon, \mathbf{r}) = -\frac{m}{(2\pi\hbar)^2}\exp[i(k_{x0}\cos\varphi_r + k_{y0}\sin\varphi_r)r]\sum_{j=1,2}\oint df \sum_{\pm}\frac{k_\pm^{(j)}}{k_\pm^{(j)} - k_\mp^{(j)}}F\left(k_\pm^{(j)}, r\cos(f - \varphi_r)\right), \epsilon \in \mathbb{C}, \quad (22)$$

$$\left\{ \begin{array}{c} g_x(\epsilon, \mathbf{r}) \\ g_y(\epsilon, \mathbf{r}) \end{array} \right\} = -\frac{m}{(2\pi\hbar)^2}\exp[i(k_{x0}\cos\varphi_r + k_{y0}\sin\varphi_r)r]\sum_{j=1,2}\oint df \left\{ \begin{array}{c} \cos\theta_j(f) \\ \sin\theta_j(f) \end{array} \right\} \sum_{\pm}\frac{k_\pm^{(j)}}{k_\pm^{(j)} - k_\mp^{(j)}}F\left(k_\pm^{(j)}, r\cos(f - \varphi_r)\right), \epsilon \in \mathbb{C}, \quad (23)$$

where $\hat{k} = k_\pm^{(1, 2)}(\epsilon)$ are the roots of the equations

$$\epsilon_{1,2}(\tilde{k}, f) = \epsilon, \quad (24)$$

$$\hat{k}_\pm^{(1,2)} = \lambda^{(1,2)} \pm \sqrt{(\lambda^{(1,2)})^2 + \frac{2m(\epsilon - \epsilon_0)}{\hbar^2}}, \quad (25)$$

$$F(k, r) = \int_0^\infty \frac{d\tilde{k}}{\tilde{k} - k}e^{i\tilde{k}r} = e^{ikr}\left[-Ci(-k|r|) + iSi(kr) + \frac{i\pi}{2}\text{sgn}\,r\right],$$
$$r \in \mathbb{R}, \ k \in \mathbb{C}, \quad (26)$$

the angle $\varphi_r$ in formulas (22) and (23) determines the direction of the vector $\mathbf{r} = r(\cos\varphi_r, \sin\varphi_r, 0)$.

In the specific case of equality of the constants of SOI $\alpha = \beta$ and the directions of the magnetic field along the axis $y = -x$ and $\mathbf{h} = h/\sqrt{2}\,(-1, 1, 0)$, the integrals in formulas (13) and (14) may be expressed through the Bessel functions:

$$g_0(\epsilon + i0, \mathbf{r}) = \frac{1}{2}(G_+(\epsilon, \mathbf{r}) + G_-(\epsilon, \mathbf{r})), \quad \epsilon \in \mathbb{R}, \quad (27)$$

$$\begin{Bmatrix} g_x(\epsilon + i0, \mathbf{r}) \\ g_y(\epsilon + i0, \mathbf{r}) \end{Bmatrix} = \pm \frac{1}{2\sqrt{2}}(G_+(\epsilon, \mathbf{r}) - G_-(\epsilon, \mathbf{r})), \quad \epsilon \in \mathbb{R}, \quad (28)$$

where

$$G_\pm(\epsilon, \mathbf{r}) = \exp\left(\pm i\frac{\sqrt{2m}\alpha}{\hbar^2}(x+y)\right) G_{2D}^R\left(\epsilon + \frac{\sqrt{2m}\alpha}{\hbar^2} \mp h, r\right), \quad (29)$$

and $G_{2D}^R(\epsilon, r)$ is the retarded Green function of free 2D electrons,

$$G_{2D}^R(\epsilon, r) = -\frac{m}{2\hbar^2}\begin{cases} iH_0^{(1)}(\sqrt{2m\epsilon}|r|/\hbar); & \epsilon > 0 \\ \frac{2}{\pi}K_0(\sqrt{2m|\epsilon|}|r|/\hbar); & \epsilon < 0 \end{cases}, \quad (30)$$

$H_0^{(n)}(x)$ is the Hankel function, and $K_0(x)$ is the modified Bessel function of the second kind.

### 3.2. Asymptotic formulas

For large $r$, the stationary phase method[39] makes it possible to derive very simple asymptotic expressions of formulas (22) and (23). Then the stationary phase $f = f_{st}^{(j)}$ should be determined from the equation

$$\frac{d}{df}\left(k_\pm^j \cos(f - \varphi_r)\right)\big|_{f=f_{st}^{(j)}} = 0, \quad (31)$$

in which the $k_\pm^{(j)}$ are the positive real solutions (25) of Eq. (24) for $\epsilon \in \mathbb{R}$.

As a result of standard calculations,[38] we obtain the following asymptotes of the components of the Green function in (12)

$$g_0(\epsilon + i0, \mathbf{r}) \simeq -\frac{i}{2\sqrt{2\pi}} \sum_{j=1,2} \sum_s \frac{1}{\hbar v^{(j)}\sqrt{|K_j|r}} \times \exp\left[iS_j r - \frac{i\pi}{4}\text{sgn}\,K_j\right]\big|_{f=f_{st}^{(j)}}, \quad \epsilon \in \mathbb{R}, \quad (32)$$

$$\begin{Bmatrix} g_x(\epsilon + i0, \mathbf{r}) \\ g_y(\epsilon + i0, \mathbf{r}) \end{Bmatrix} \simeq -\frac{i}{2\sqrt{2\pi}} \sum_{j=1,2} \sum_s \begin{Bmatrix} \cos\theta_j \\ \sin\theta_j \end{Bmatrix} \frac{1}{\hbar v^{(j)}\sqrt{|K_j|r}} \exp\left[iS_j r - \frac{i\pi}{4}\text{sign}\,K_j\right]\big|_{f=f_{st}^{(j)}}, \quad \epsilon \in \mathbb{R}, \quad (33)$$

$$S_j(f, \varphi_r) = k_\pm^{(j)}(f)\cos(f - \varphi_r), \quad (34)$$

which are valid for $S_j r \gg 1$. All functions must be calculated at the points of the stationary phase $f = f_{st}^{(j)}$. The summation on $s$ takes account of the possibility of the existence, on a non-convex isoenergy contour belonging to a branch of the spectrum $\epsilon_2(\mathbf{k})$, of several points of the stationary phase corresponding to the given direction of the vector $\mathbf{r}$. In formulas (32) and (33), $K_{1,2}(f) \neq 0$ is the curvature of the isoenergy curve $\epsilon_{1,2}(f) = \epsilon$.

$$K_j(f) = \frac{k_\pm^{(j)}(f)^2 + 2\dot{k}_\pm^{(j)}(f)^2 - k_\pm^{(j)}(f)\ddot{k}_\pm^{(j)}(f)}{\left(k_\pm^{(j)}(f) + \dot{k}_\pm^{(j)}(f)^2\right)^{3/2}}, \quad (35)$$

and $vj \neq 0$ is the absolute value of the velocity of an electron $\mathbf{v}^{(j)} = \nabla_\mathbf{k}\,\epsilon_j/\hbar$. Hereinafter, a dot over a function signifies differentiation on the angle $f$ of the direction of a wave vector in the displaced coordinates (16). Solutions of Eq. (31) satisfying the inequality $S_j(f)$ [see formula (34)] correspond to the condition of parallelism of vectors $\mathbf{r}$ and $\mathbf{v}^{(j)}$ (see also Ref. 43)

$$\mathbf{r}\mathbf{n}_v^{(j)}\big|_{f=f_{st}^{(j)}} = r; \quad \mathbf{n}_v^{(j)}(f) = \frac{\mathbf{v}^{(j)}}{|\mathbf{v}^{(j)}|}$$

$$= \mp\left(-\frac{k_\pm^{(j)}\sin f + \dot{k}_\pm^{(j)}\cos f}{\sqrt{k_\pm^{(j)2} + \dot{k}_\pm^{(j)2}}}, \frac{k_\pm^{(j)}\cos f - \dot{k}_\pm^{(j)}\sin f}{\sqrt{k_\pm^{(j)2} + \dot{k}_\pm^{(j)2}}}\right). \quad (36)$$

The values of the phase $\mathbf{kr}$ of rapidly oscillating functions as $r \to \infty$ in the integrals in (13) and (14) can be interpreted[43] in terms of the support function[40] of the isoenergy contour $\epsilon_{1,2}(\mathbf{k}) = \epsilon$.

$$S_j(\mathbf{k}) = \mathbf{k}\mathbf{n}_v^{(j)}(\mathbf{k}); \quad \mathbf{k} \in \epsilon_i(\mathbf{k}) = \epsilon, \quad (37)$$

knowing which, it is possible to restore the contour and find its curvature at any point.

### 4. Friedel oscillations

#### 4.1. Fermi contours

It was shown in Refs. 41–43 that the geometry of the constant phase lines of the FO oscillations of the LDS and their period depend on the local geometry of the Fermi surface. Therefore, in this section we will introduce some information that will be needed later regarding the energy spectrum of the system being studied.

In the case of 2D electron gas with R-D SOI placed in a parallel magnetic field, the energy spectrum contains two branches $\epsilon_{1,2}(\mathbf{k}) \neq \epsilon_{1,2}(-\mathbf{k})$ (2), not possessing central symmetry. The surface $\epsilon = \epsilon_1(\mathbf{k})$ is always convex, at the same time that the surface $\epsilon = \epsilon_2(\mathbf{k})$ for a defined region of values of SOI constants and magnetic field contains saddle points and areas of negative Gaussian curvature (see for example Refs. 34 and 35. There exists a critical value of the

magnetic field

$$h_c = \frac{(\alpha^2 - \beta^2)^2}{\sqrt{\alpha^4 + 6\alpha^2\beta^2 + \beta^4 + 4\alpha\beta(\alpha^2 + \beta^2)\sin 2\varphi_h}}, \quad (38)$$

for values less than which, for $h < h_{c2}$ the function $\epsilon = \epsilon_1(\mathbf{k})$ has no extrema and takes the smallest value $\epsilon = \epsilon_0$ (18) at the point of contact of the branches of the spectrum, and for $h > h_c$, $\epsilon = \epsilon_1(\mathbf{k})$ has an absolute minimum.

In the coordinate system of (16), the equations of Fermi contours $k(\epsilon_F, f) = k_\pm^{(1,2)}$ are defined by the formulas (25) for $\epsilon = \epsilon_F$ in the field of values of parameters for which $k_\pm^{(1,2)} \geq 0$. In the case when the inequality $\epsilon_F > \epsilon$ is satisfied, the functions $k_+^{(1,2)}(\epsilon_F, f) > 0$ for any values of $f$ while at the same time the roots $k_-^{(1,2)}(\epsilon_F, f) < 0$ are always negative. To each branch of the spectrum there corresponds one Fermi contour $k(\epsilon_F, f) = k_+^{(1,2)}$. For $\epsilon_F < \epsilon_0$ the real roots of Eq. (24) exist when the inequality is satisfied

$$\frac{2m(\epsilon_0 - \epsilon_F)}{\hbar^2} \leq (\lambda^{(j)})^2,$$

and both roots $k_\pm^{(j)}$ take positive values in some interval of angles for which $\lambda_j \geq 0$. In this case the Fermi contours do not cover the coordinate origin $(k_{x0}, k_{y0})$ (17), and the two points $k = k_\pm^{(j)}$ on the same Fermi contour correspond to the same direction of the angle $f$. Due to the dependence of the energy at the point of contact of the branches of the spectrum $\epsilon_0$ (18) and the functions $\lambda^{(j)}(f)$ (20) on $\mathbf{h}$, it is possible by varying the magnitude and direction of the magnetic field to change smoothly the energy spectrum and therefore the geometry of isoenergy contours.

### 4.2. LDS oscillations

Using expressions (8) and (10), we will present LDS in the form of the sum of two components:

$$\rho(\epsilon_F, \mathbf{r}) = \rho_0(\epsilon_F) + \Delta\rho(\epsilon_F, \mathbf{r}), \quad (39)$$

where $\rho_0(\epsilon_F)$ is the density of states of two-dimensional degenerate gas from R-D SOI in the absence of defects:[34,38]

$$\rho_0(\epsilon_F) = \begin{cases} \dfrac{m}{\pi\hbar^2}; & \epsilon_F \geq \epsilon_0 \\ \dfrac{m}{2\pi^2\hbar^2} \sum_{j=1,2} \oint df \dfrac{\lambda^{(j)}}{\sqrt{\xi^{(j)}}} \Theta(\lambda^{(j)})\Theta(\xi^{(j)}); & \epsilon_F \leq \epsilon_0, \end{cases} \quad (40)$$

$$\xi^{(1,2)} = (\lambda^{(1,2)})^2 + \frac{2m(\epsilon_F - \epsilon_0)}{\hbar^2}. \quad (41)$$

The part of LDS $\Delta\rho(\epsilon_F, \mathbf{r})$ that depends on scattering describes the FO. Using decomposition (12) of the Green function over Pauli matrices, we derive

$$\Delta\rho(\mathbf{r}) = -\frac{2}{\pi} \text{Im}\{\gamma[g_0(\mathbf{r})g_0(-\mathbf{r}) + g_x(\mathbf{r})g_x(-\mathbf{r}) + g_y(\mathbf{r})g_y(-\mathbf{r})] + J_x[g_0(\mathbf{r})g_x(-\mathbf{r}) + g_x(\mathbf{r})g_0(-\mathbf{r})] \\ + J_y[g_0(\mathbf{r})g_y(-\mathbf{r}) + g_y(\mathbf{r})g_0(-\mathbf{r})] + iJ_z[g_y(\mathbf{r})g_x(-\mathbf{r}) - g_x(\mathbf{r})g_y(-\mathbf{r})]\}. \quad (42)$$

The lack of a center of inversion of the electron scattering law (2) results in the appearance in $\Delta\rho$ in (42) of a component proportional to the components $J_{x,y,z}$ of the magnetic moment of the defect, which goes to zero for $h = 0$. Using asymptotic expressions (32) and (33) for the Green functions at large distances from the defect, we derive

$$\Delta\rho(\mathbf{r}) = -\sum_{i,j=1,2} \sum_s Q_{ij} \left\{ \left( 2\gamma \cos^2\left[\frac{\theta_i - \overline{\theta}_j}{2}\right] + J\sin\theta_J \left(\cos(\varphi_J - \varphi_i) + \cos(\varphi_J - \overline{\theta}_j)\right)\right) \right. \\ \left. \times \sin\left((S_i + \overline{S}_j)r + \phi_{ij}\right) - J\cos\theta_J \sin(\theta_i - \overline{\theta}_j)\cos\left((S_i + \overline{S}_j)r + \phi_{ij}\right) \right\}, \quad (43)$$

$$Q_{ik} = \frac{1}{2\pi^2\hbar v^{(i)}\overline{v}^{(j)}|K_i\overline{K}_j|r}; \quad \phi_{ij} = -\frac{\pi}{4}(\text{sign}\, K_i + \text{sign}\, \overline{K}_j). \quad (44)$$

The angles $\varphi_J$ and $\theta_J$ specify the direction of the vector of the magnetic moment of the defect.

$$\mathbf{J} = J(\cos\varphi_J \sin\theta_J, \sin\varphi_J \sin\theta_J, \cos\theta_J). \quad (45)$$

The line over a function signifies that its value is taken at the point $\mathbf{k} = \hat{\mathbf{k}}_{st}^{(1,2)}$ for which the velocity $\mathbf{v}^{(1,2)}$ of an electron is directed opposite to the direction of the vector $\mathbf{r}$, and $\mathbf{n}_v^{(1,2)}(\hat{\mathbf{k}}_{st}^{(1,2)}) = -\mathbf{n}_v^{(1,2)}(\mathbf{k}_{st}^{(1,2)})$. Hence, each of the components in the sum (43) takes account of the contribution to LDS of electron back-scatter with a transition between two points with opposite direction of the velocity on the same ($i = j$) or different ($i \neq j$) Fermi contours.

The asymptotic formula in (43) makes it possible to easily interpret the reason for the occurrence of a magnetic contribution to the density of states: due to the connection between the directions of spin and the wave vector, the magnetic scattering that results in spin flip at some angle $\Delta\theta_i = \theta_i$

– $\bar{\theta}_j \neq 0, \pm\pi$ and that corresponds to the change of velocity of an electron to its opposite, changes the flux of electrons propagated in the opposite direction.

### 4.3. LDM oscillations

We present an expression for LDM $\mathbf{m}(\mathbf{r})$ in the form of a sum not dependent on the coordinates of the component $\mathbf{m}_0$ and the oscillatory additive $\Delta\mathbf{m}(\mathbf{r})$

$$\mathbf{m}(\mathbf{r}) = \mathbf{m}_0 + \Delta\mathbf{m}(\mathbf{r}). \tag{46}$$

It is clear that due to the common relationships (12) and (9), the component of the density of magnetization $m_{0z} = 0$. However, the components $m_{0x, y}$ may be distinct from zero at the Fermi energy $\epsilon_F \leq \epsilon_0$:

$$m_{x,y} = \begin{cases} 0, & \epsilon_F \geq \epsilon_0, \\ \dfrac{m}{2\pi^2\hbar^2} \displaystyle\sum_{j=1,2} \oint df \begin{Bmatrix} \cos\theta_i \\ \sin\theta_i \end{Bmatrix} \dfrac{\lambda^{(j)}}{\sqrt{\xi^{(j)}}} \Theta(\lambda^{(j)})\Theta(\xi^{(j)}), & \epsilon_F \leq \epsilon_0. \end{cases} \tag{47}$$

Using a representation of the Green function in the form of decomposition on Pauli matrices (12), we will write expressions for the component of the vector $\Delta\mathbf{m}(\mathbf{r})$ in the following form:

$$\Delta m_x(\mathbf{r}) = -\frac{2}{\pi}\text{Im}\{J_x[g_0(\mathbf{r})g_0(-\mathbf{r}) + g_x(\mathbf{r})g_x(-\mathbf{r}) - g_y(\mathbf{r})g_y(-\mathbf{r})] + J_y[g_x(\mathbf{r})g_y(-\mathbf{r}) + g_y(\mathbf{r})g_x(-\mathbf{r})]$$
$$- iJ_z[g_0(\mathbf{r})g_y(-\mathbf{r}) + g_y(\mathbf{r})g_0(-\mathbf{r})] + \gamma[g_0(\mathbf{r})g_x(-\mathbf{r}) + g_x(\mathbf{r})g_0(-\mathbf{r})]\}, \tag{48}$$

$$\Delta m_y(\mathbf{r}) = -\frac{2}{\pi}\text{Im}\{J_y[g_0(\mathbf{r})g_0(-\mathbf{r}) + g_x(\mathbf{r})g_x(-\mathbf{r}) - g_y(\mathbf{r})g_y(-\mathbf{r})] + J_x[g_y(\mathbf{r})g_x(-\mathbf{r}) + g_x(\mathbf{r})g_y(-\mathbf{r})]$$
$$- iJ_z[g_0(\mathbf{r})g_x(-\mathbf{r}) + g_x(\mathbf{r})g_0(-\mathbf{r})] + \gamma[g_0(\mathbf{r})g_y(-\mathbf{r}) + g_y(\mathbf{r})g_0(-\mathbf{r})]\}, \tag{49}$$

$$\Delta m_y(\mathbf{r}) = -\frac{2}{\pi}\text{Im}\{J_z[g_0(\mathbf{r})g_0(-\mathbf{r}) + g_x(\mathbf{r})g_x(-\mathbf{r}) - g_y(\mathbf{r})g_y(-\mathbf{r})] + iJ_x[g_0(\mathbf{r})g_y(-\mathbf{r}) + g_y(\mathbf{r})g_0(-\mathbf{r})]$$
$$- iJ_y[g_0(\mathbf{r})g_x(-\mathbf{r}) - g_x(\mathbf{r})g_0(-\mathbf{r})] + i\gamma[g_x(\mathbf{r})g_y(-\mathbf{r}) - g_y(\mathbf{r})g_x(-\mathbf{r})]\}. \tag{50}$$

Using the derived formulas (48)–(50) and asymptotic expressions (32)–(33) for the component of the Green function (12) for large $r$, we find the LDM components that are oscillatory with distance from the defect:

$$\Delta m_x(\mathbf{r}) = -\sum_{i,j=1,2}\sum_s Q_{ij}\left\{\left[J\sin\theta_j\left(\cos\varphi_J\cos^2\left(\frac{\theta_i+\bar{\theta}_j}{2}\right)\right) + \sin\varphi_J\sin(\theta_i+\bar{\theta}_j)\right] + J\cos\theta_J(\sin\theta_i - \sin\bar{\theta}_j)\right.$$
$$\left. \times \cos((S_i+\bar{S}_j)r + \phi_{ij}) + \gamma(\cos\theta_i + \sin\theta_i + \cos\bar{\theta}_j + \sin\bar{\theta}_j)] \times \sin((S_i+\bar{S}_j)r + \phi_{ij})\right\}, \tag{51}$$

$$\Delta m_y(\mathbf{r}) = -\sum_{i,j=1,2}\sum_s Q_{ij}\left\{\left[J\sin\theta_j\left(2\sin\varphi_J\sin^2\left(\frac{\theta_i+\bar{\theta}_j}{2}\right)\right) + \cos\varphi_J\sin(\theta_i+\bar{\theta}_j)\right] - J\cos\theta_J(\sin\theta_i - \sin\bar{\theta}_j)\right.$$
$$\left. \times \cos((S_i+\bar{S}_j)r + \phi_{ij}) + \gamma(\cos\theta_i + \sin\theta_i + \cos\bar{\theta}_j + \sin\bar{\theta}_j)]\sin((S_i+\bar{S}_j)r + \phi_{ij})\right\}, \tag{52}$$

$$\Delta m_z(\mathbf{r}) = -\sum_{i,j=1,2}\sum_s Q_{ij}\left\{2J\cos\theta_j\sin^2\left(\frac{\theta_i+\bar{\theta}_j}{2}\right) + \sin((S_i+\bar{S}_j)r + \phi_{ij}) - [J\sin\theta_J(\sin\varphi_J(\sin\theta_i - \sin\bar{\theta}_j)\right.$$
$$\left. - \cos\varphi_J(\cos\theta_i - \cos\bar{\theta}_j)) + \gamma\sin(\theta_i - \bar{\theta}_j)]\cos((S_i+\bar{S}_j)r + \phi_{ij})\right\}. \tag{53}$$

The derived results (51)–(53) define the dependence of the FO of the LDM on an external magnetic field. As in the case of LDS, the non-magnetic contribution to LDM can be easily interpreted on the basis of the asymptotic formulas (51)–(53): back-scatter by a non-magnetic defect is accompanied by a transition to a state with the spin direction turned relative to the initial at an angle $\Delta\theta_i = \theta_i - \bar{\theta}_j \neq 0, \pm\pi$, which results in a change of the local density of magnetization around the defect.

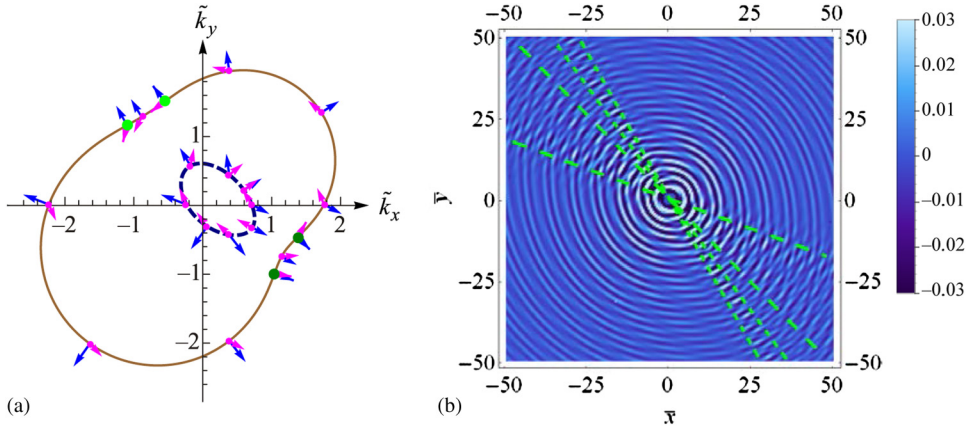

Fig. 1. (a) Typical form of Fermi contours for $\epsilon_F > \epsilon_0$ in a parallel magnetic field. (b) Oscillatory part of LDS $\Delta\bar{\rho}(\bar{x}, \bar{y})$ for scattering by a non-magnetic defect ($J = 0$). The following values of parameters are used: $\bar{\alpha} = 0.7$, $\bar{\beta} = 0.3$, $\bar{h} = 0.6$, $\varphi_h = 2.0$.

### 4.4. The special case $\alpha = \beta$ and $h_x = -h_y$. Analytical solution

In the specific case that $\alpha = \beta$ and $\mathbf{h} = h\sqrt{2}\,(-1,\,1,\,0)$, two branches of the spectrum are crossed along the parabola

$$\epsilon = \frac{\hbar^2 k_{y1}^2}{2m} + \frac{\hbar^2 h^2}{8m\alpha^2}, \quad k_{x1} = \frac{k_x + k_y}{\sqrt{2}} = \frac{h}{2\alpha}, \quad k_{y1} = \frac{k_x - k_y}{\sqrt{2}}. \tag{54}$$

For $\epsilon_F > \hbar^2 h^2 / 8m\alpha^2$, the Fermi contours have two common points and each of them consists of two arcs of circles of radius $k^{(\pm)} = \sqrt{(2m\epsilon_\pm)}/\hbar$ (see Fig. 5):

$$k^{(\pm)} = \sqrt{2m|\epsilon_\pm|}/\hbar, \quad \epsilon_\pm = \epsilon_F + \frac{2m\alpha^2}{\hbar^2} \mp h. \tag{55}$$

The spin directions on each of the arcs that comprise one contour are opposite: $\theta_+ = 3\pi/4$ or $\theta_- = -\pi/4$, and the arcs with the identical spin directions $\theta_\pm$ form complete circles [see Fig. 5(a)].

Using the expressions for the Green functions in (27) and (28), we derive the following expression for the oscillatory part of the LDS:

$$\Delta\rho(r) = \left(\gamma + \frac{J_y - J_x}{2\sqrt{2}}\right) R_+ + \left(\gamma - \frac{J_y - J_x}{2\sqrt{2}}\right) R_-, \tag{56}$$

where

$$R_\pm(r) = \frac{m^2}{4\pi\hbar^4} J_0(k^{(\pm)}r) Y_0(k^{(\pm)}r) \Theta(\epsilon_+), \tag{57}$$

where $J_0(x)$ and $Y_0(x)$ are the Bessel functions of the first and second kind, respectively. For large values of the arguments $k^{(\pm)} r \gg 1$, we have

$$R_\pm(r) \simeq -\frac{m^2}{4\pi^2\hbar^4 k^{(\pm)} r} \cos(2k^{(\pm)}r). \tag{58}$$

The components of the oscillatory part of the LDM, $\Delta\mathbf{m}(\mathbf{r})$, are defined by the following expressions:

$$\Delta m_x(\mathbf{r}) = \frac{\gamma}{\sqrt{2}} R_+ + \left(\frac{\gamma}{\sqrt{2}} - \frac{J_y - J_x}{2}\right) R_- - \frac{J_z}{2} R, \tag{59}$$

$$\Delta m_y(\mathbf{r}) = \left(\frac{\gamma}{\sqrt{2}} + \frac{J_y - J_x}{2}\right) R_+ - \frac{\gamma}{\sqrt{2}} R_- - \frac{J_z}{2} R, \tag{60}$$

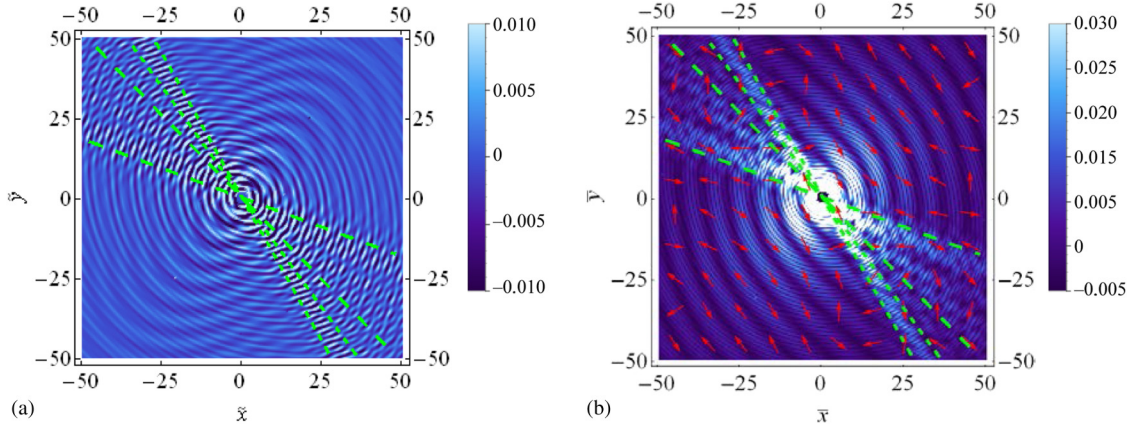

Fig. 2. FO of the LDM for scattering by a a non-magnetic defect with $\epsilon_F > \epsilon_0$. (a) Distribution of the component $\Delta\bar{m}_z$ normal to the plane. (b) Distribution of the absolute value $\sqrt{\Delta\bar{m}_x^2 + \Delta\bar{m}_y^2}$ of the LDM component in the plane. Arrows indicate the direction of the vector $\Delta\bar{\mathbf{m}} = (\Delta\bar{m}_x, \Delta\bar{m}_y)$. Values of the parameters coincide with those given in Fig. 1.

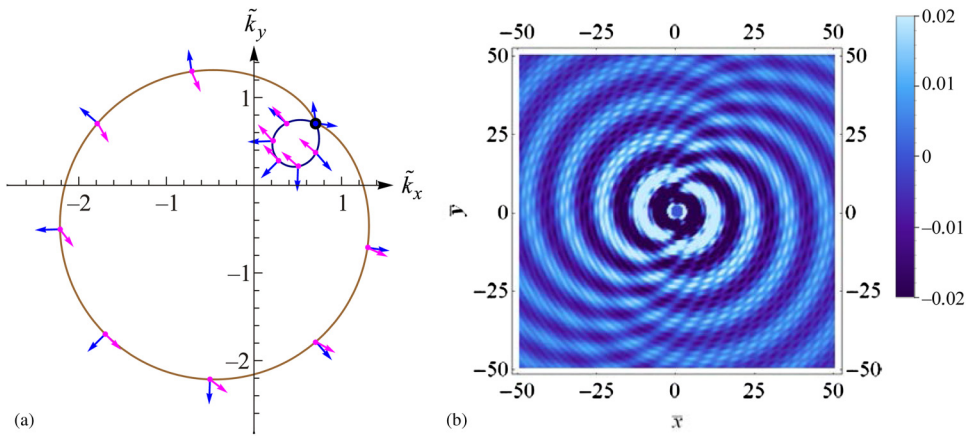

Fig. 3. (a) Fermi contours for $\epsilon_0 = \epsilon_F$ (18), $h > h_c$ (38), $\varphi_h = 3\pi/4$. The black point shows $\mathbf{k}_0$, the point of contact of the branch of the range of (17). (b) LDS for scattering by magnetic defect $\mathbf{J} = (J, 0, 0)$ and $\gamma = 0$ (7). The following values of parameters are used: $\bar{\alpha} = 0.5$, $\bar{\beta} = 0.2$, $\bar{h} = 1.4$.

$$\Delta m_z(\mathbf{r}) = \frac{J_z}{2}(R_+ - R_-) \frac{J_y - J_x}{2\sqrt{2}} R, \qquad (61)$$

where the designations $R_\pm(r)$ are defined by expression (57) and

$$R(\mathbf{r}) = \frac{m^2}{2\pi\hbar^4} \sin\left[\frac{2\sqrt{2}m\alpha}{\hbar^2}(x+y)\right] \left\{ J_0(k^{(+)}r)Y_0(k^{(-)}r) + J_0(k^{(-)}r)Y_0(k^{(+)}r) \right\} \Theta(\epsilon_+)\Theta(\epsilon_-), \qquad (62)$$

$$R(\mathbf{r}) \simeq -\frac{4m^2}{\pi^2\hbar^4 r\sqrt{k^{(+)}k^{(-)}}} \left\{ \sin\left[\left(k^{(+)} + k^{(-)} + \frac{4m\alpha}{\hbar^2}\sin\left(\varphi_r + \frac{\pi}{4}\right)\right)r\right] - \sin\left[\left(k^{(+)} + k^{(-)} + \frac{4m\alpha}{\hbar^2}\sin\left(\varphi_r + \frac{\pi}{4}\right)\right)r\right] \right\}; \quad k^{(\pm)}r \gg 1. \quad (63)$$

In the case being studied, the FO of the LDS (56) are isotropic, and at large distances from the defect contain two harmonics with periods $\Delta r = \pi/k^{(\pm)}$, associated with back-scatter between states belonging to different Fermi contours, while at the same time in LDM oscillation (59)–(61), a contribution is made by the transitions between states of the same Fermi contour that result in the appearance of harmonics of the FO, with periods

$$\Delta r = \frac{2\pi}{k^{(+)} + k^{(-)} \pm \frac{4m\alpha}{\hbar^2}\sin\left(\varphi_r + \frac{\pi}{4}\right)}, \qquad (64)$$

depending on the direction $\varphi_r$ in the coordinate space. It follows from formula (63) that the lines of constant phase for oscillations of the LDM with the period in (64) are ellipses (for $k^{(+)} + k^{(-)} > 4m\alpha/\hbar^2$) or hyperbolas (for $k^{(+)} + k^{(-)} < 4m\alpha/\hbar^2$), and the point $r = 0$ coincides with one of the foci.

## 5. Discussion of results

The effect of a parallel magnetic field the FO of a 2D electron gas with Rashba-Dresselhaus SOI results from two main causes. First, the magnetic field $\mathbf{B}$ breaks the central symmetry of the Fermi contours and changes their local geometry. The back-scatter of electrons, which provides the chief contribution to the FO, corresponds to transitions between states with opposite direction of velocity which in turn depends on the field **B**. As a result, a change to the magnitude and direction of the vector **B** changes both the period of the FO [as a consequence of the change to the magnitude of the wave vector corresponding to a point of a stationary phase, see (34)] and their amplitude [as a consequence of the change of the curvature of the Fermi contour, see (35)]. The second main circumstance is the change of electron spin under the effect of the direction of the field [see (6)]. Since with SOI the spin of an electron and its wave vector are interconnected, the matrix elements of transitions between two quantum states with back-scatter depend on the direction of the spins before and after scattering. For **B** = 0, replacement of the direction of a wave vector by its opposite results in a spin-flip for states on the same Fermi contour, and to preserving its direction upon transition to another Fermi contour. As a result, due to the selection rules for spin, there are no components in the FO of the LDS that depend on the magnetic defect moment, and there are naturally no FO of the LDM during scattering by a non-magnetic defect (see Refs. 23, 26, and 33). In a parallel magnetic field during back-scatter, states corresponding to a spin-flip to some angle depending on the field **B** are permissible. As a result, new harmonics appear in the FO of the LDS, whose periods depend only on the characteristics of one of the Fermi contours, as well as components proportional to the magnetic defect moment $J$. For the same reason, the FO of the LDM contain harmonics proportional to the constant $\gamma$ of the potential interaction of an electron with a defect. Note that the listed conclusions remain valid in the presence of only one SOI type (Rashba or Dresselhaus), and all our analytical results make it possible to set one of the constants of SOI equal to zero. Asymptotic formulas (43) and (51)–(53) provide a complete qualitative description of all harmonics of the FO of the LDM and LDS, and the dependences of their periods and amplitudes on the field **B**.

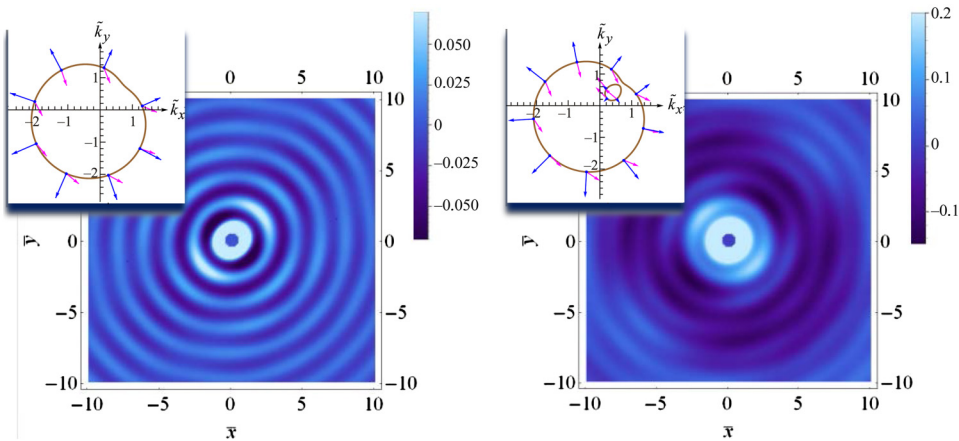

Fig. 4. FO of the LDS $\Delta\bar{\rho}(\bar{\mathbf{r}})$ with scattering by a magnetic defect, $\mathbf{J} = (1, -1, 0)J_0/\sqrt{2}$ and $\gamma = 0$, near the value of the magnetic field $\bar{h}_{\min 1} = 1.36$, where the Fermi level passes through the point of the minimum of the branch of the range $\epsilon_1$. (a) $\bar{h}_{\min 1} < \bar{h} = 1.4$. (b) $\bar{h}_{\min 1} > \bar{h} = 1.28$. The following values of parameters are used: $\bar{\alpha} = 0.5$, $\bar{\beta} = 0.1$, $\varphi_h = 3\pi/4$.

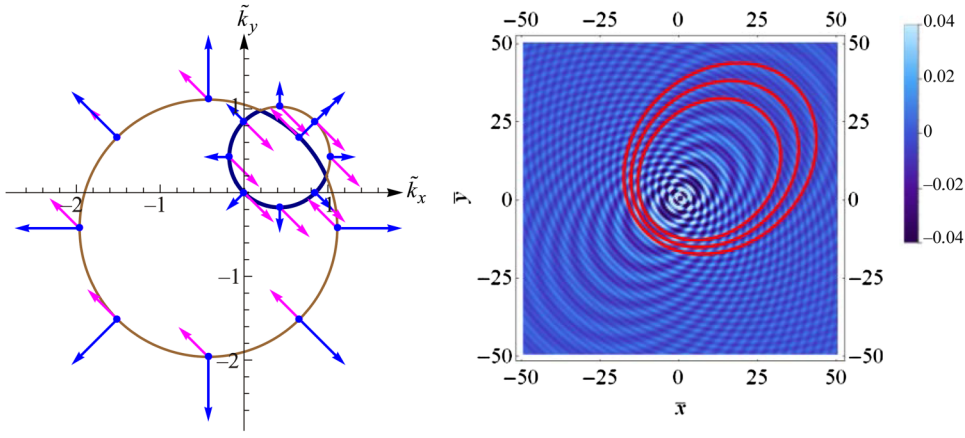

Fig. 5. FO of the LDS $\bar{\rho}$ in the case of equality of constants of R-D SOI for scattering by a non-magnetic defect, $\gamma \neq 0$ and $\mathbf{J} = 0$. (a) Fermi contours, (b) oscillations of the z-component of the LDM, $\bar{m}_z$, for scattering by a magnetic defect, $\gamma = 0$ and $\mathbf{J} = (0, 0, J)$. The following values of parameters were used in constructing the graphs: $\bar{\alpha} = \bar{\beta} = 0.3$, $\bar{h} = 1$, $\varphi_h = 3\pi/4$.

The spatial distributions of local densities of states and magnetization in Figs. 1–5, obtained by means of the general expressions (42) and (48)–(50), are only several specific examples illustrating the variety of the nature of the anisotropy of Friedel oscillations in the system under study. We use the following dimensionless values in the creation of the diagrams:

$$\bar{\alpha} = \frac{m\alpha}{\hbar^2 k_F}, \quad \bar{\beta} = \frac{m\beta}{\hbar^2 k_F}, \quad \bar{h} = \frac{h}{\epsilon_F} = \frac{2mh}{\hbar^2 k_F^2},$$
$$\bar{k} = \frac{k}{k_F}, \quad \bar{r} = k_F r, \quad \bar{\epsilon} = \frac{\epsilon}{\epsilon_F}, \quad (65)$$

$$\Delta\bar{\rho}(\bar{r}) = \left(\frac{\pi^2 \hbar^4}{m^2(\gamma + J)}\right)\Delta\rho(r); \quad \Delta\bar{\mathbf{m}}(\bar{r}) = \left(\frac{\pi^2 \hbar^4}{m^2(\gamma + J)}\right)\Delta\mathbf{m}(r), \quad (66)$$

where $k_F = \sqrt{(2m\epsilon_F)}/\hbar$. The arrows placed on the diagrams on the Fermi contours show the direction of the vector of velocity (arrows directed perpendicular to the line of the Fermi contour) and the direction of electron spin.

Fig. 1(a) demonstrates a violation of the central symmetry of the Fermi contours and a change in direction of the spins under the effect of a parallel magnetic field under conditions of R-D SOI. With scattering by a non-magnetic defect, the FO of the LDS $\Delta\bar{\rho}(\bar{r})$ [Fig. 1(b)] preserve the central symmetry, but no longer have symmetry relative to axes $x = y$ and $x = -y$, which exists in the absence of field $\mathbf{B} = 0$.[33] The dashed lines in Fig. 1(b) show the directions of the maximum amplitude of the FO that coincide with the direction of velocity at the inflection points of Fermi contour belonging to branch $\epsilon_2$. Each two lines (with identical length of a stroke) limit the "fan" of the directions, in which the FO have more than two harmonics. In the zero field, both "fans" coincide.

Fig. 2 illustrates the effect of the emergence of a nonuniform distribution of the density of magnetization existing only in a magnetic field with scattering by a non-magnetic defect. It is interesting to note that potential scattering leads to nonzero density of magnetization not only in the plane 2D of electrons [Fig. 1, Eq. (6)] in which the vector **B** lies, but also in the direction perpendicular to this plane [Fig. 1(a)].

In Fig. 3, the FO of the LDS caused by magnetic scattering for the special case when the Fermi energy coincides with the energy $\epsilon_0$ at the point of a contact of the branches of the range in (18), and a magnetic field greater than the critical $h_c$ (38), i.e., a branch of the range $\epsilon = \epsilon_1(\mathbf{k})$ are given has an absolute minimum with energy $\epsilon_1^{\min} < \epsilon_0$.

Figure 4 visually demonstrates the essential change in the nature of the FO of the LDS associated with magnetic

scattering in a narrow interval of magnetic fields near the value

$$h = h_{\min}^{(1)} = \frac{m}{2\hbar^2}(\alpha + \beta)^2 + \epsilon_F, \varphi_h = \frac{3\pi}{4}, \qquad (67)$$

corresponding to the decreases in the minimum of the branch of range $\epsilon_1$ with Fermi level $\epsilon_F$. For $h > h_{\min}^{(1)}$ there is only one Fermi contour belonging to branch $\epsilon_2$ [inset in Fig. 4(a)] and defining the unique harmonic of the FO in Fig. 4(a). For magnetic fields lower than the minimum field $h < h_{\min}^{(1)}$, a second Fermi contour occurs [inset in Fig. 4(b)], belonging to branch $\epsilon_1$ (a 2D analog of the topological Lipshitz transition with the appearance of a new cavity of the Fermi surface), which results in the emergence of a second harmonic of FO with greater period and amplitude, as in Fig. 4(b). The distributions of the LDS are constructed for the case where $h > h_c$ (38) and $\epsilon_F < \epsilon_0$ (18).

Figure 5 pertains to the special case $\alpha = \beta$ and $\mathbf{h} = h/\sqrt{2}$ (−1, 1, 0), examined in section 5.3. For the chosen values of parameters, the Fermi contours have two common points, i.e., $\epsilon_F > \hbar^2 h^2 / 8m\alpha^2$ [Fig. 5(a)]. According to the asymptotic formula (63), there are visible in Fig. 5 ellipses of lines of constant phase in the spatial distribution of the LDS component $\bar{m}_z$ around the magnetic defect.

## 6. Conclusion

We investigated the effect of a parallel magnetic field on Friedel oscillations of the local densities of states and of magnetization in a 2D electron gas with R-D SOI that are associated with scattering by a magnetic defect. It is shown that a magnetic field breaking the central symmetry of the law of scattering results in the appearance in the FO of the LDS of harmonics caused by magnetic scattering, and the non-magnetic impurity generates FO of the LDM. The predicted effect opens up the possibility of researching magnetic scattering by means of the usual rather than spin-polarized STM. The dependence of the periods of FO on the value and direction of the vector of a magnetic field may serve as an independent method of determining the constants of SOI.

a)Email: kolesnichenko@ilt.kharkov.ua